\documentclass[preprint2]{aastex}

\usepackage{bm}

\def\la{\;
\raise0.3ex\hbox{$<$\kern-0.75em\raise-1.1ex\hbox{$\sim$}}\; }
\def\ga{\;
\raise0.3ex\hbox{$>$\kern-0.75em\raise-1.1ex\hbox{$\sim$}}\; }

\def\veps{\varepsilon}

\newcommand{\chhhoh}{{CH$_3$OH}}

\newcommand{\etal}{{et al.}}

\newcommand{\ms}{m~s$^{-1}$}
\newcommand{\kms}{km~s$^{-1}$}
\newcommand{\dmm}{$\Delta \mu/\mu$}

\slugcomment{Submitted to ApJ}

\shorttitle{$m_{\rm e}/m_{\rm p}$ constraints}
\shortauthors{Levshakov, Kozlov, \& Reimers}

\begin{document}


\title{Methanol as a tracer of fundamental constants}


\author{S. A. Levshakov}
\affil{A. F. Ioffe Physical-Technical Institute, Saint Petersburg 194021, Russia}
\email{lev@astro.ioffe.rssi.ru}

\author{M. G. Kozlov}
\affil{Petersburg Nuclear Physics Institute, Gatchina 188300, Russia}
\email{mgk@mf1309.spb.edu}

\and

\author{D. Reimers}
\affil{Hamburger Sternwarte, Universit\"at Hamburg, Gojenbergsweg 112, D-21029 Hamburg, Germany}
\email{st2e101@hs.uni-hamburg.de}



\begin{abstract}
The methanol molecule CH$_3$OH has a complex 
microwave spectrum with a large number of very
strong lines. This spectrum includes purely rotational transitions as well as
transitions with contributions of the internal degree of freedom associated
with the hindered rotation of the OH group. The latter takes place due to the
tunneling of hydrogen through the potential barriers between three equivalent
potential minima. 
Such transitions are highly sensitive to changes in
the electron-to-proton mass ratio, $\mu = m_{\rm e}/m_{\rm p}$, 
and have different responses to $\mu$-variations.
The highest 
sensitivity is found for the mixed rotation-tunneling transitions
at low frequencies. 
Observing methanol lines provides more stringent limits
on the hypothetical variation of $\mu$ than ammonia observation 
with the same velocity resolution.
We show that the best quality radio astronomical data on methanol maser lines 
constrain the variability of $\mu$ in the Milky Way at the level of
$|\Delta \mu/\mu| < 28\times10^{-9}$ ($1\sigma$) 
which is in line with the previously obtained ammonia result, 
$|\Delta \mu/\mu| < 29\times10^{-9}$ ($1\sigma$).
This estimate can be further improved if the rest frequencies of the \chhhoh\ microwave lines
will be measured more accurately.
\end{abstract}


\section{Introduction}
\label{sect-1}

The hypothetical variation of the dimensionless physical constant $\mu$~--
the electron-to-proton mass ratio~-- can be probed through the spectral observations of
certain molecular transitions which are particularly sensitive to changes in $\mu$.
The corresponding
sensitivity coefficients $Q_\mu$ of different molecular transitions relevant to astrophysics
were calculated at first for H$_2$ (Varshalovich \& Levshakov 1993), and later on for
OH (Darling 2003; Chengalur \& Kanekar 2003),
$^{15}$ND$_3$ (van Veldhoven \etal\ 2004), 
H$_2$ (Ubachs \etal\ 2007),
NH$_3$ (Flambaum \& Kozlov 2007), 
OH and CH (Kozlov 2009), 
NH$_2$D and ND$_2$H (Kozlov \etal\ 2010),
H$_3$O$^+$ (Kozlov \& Levshakov 2011),
CH$_3$OH (Jansen \etal\ 2011), and
H$_3$O$^+$, H$_2$DO$^+$, HD$_2$O$^+$, and D$_3$O$^+$ (Kozlov \etal\ 2011).
Among them ammonia, NH$_3$, 
is actively used in extragalactic and galactic observations
of dense molecular clouds. In extragalactic observations the most stringent
limits on the mass ratio,
$\Delta \mu/\mu = (\mu_{\rm space} - \mu_{\rm lab})/\mu_{\rm lab}$,
were obtained by Henkel \etal\ (2009) at redshift $z = 0.89$ and 
by Kanekar (2011) at $z = 0.69$: 
$|\Delta \mu/\mu| < 1400$ ppb $(3\sigma)$, and $|\Delta \mu/\mu| < 400$ ppb $(3\sigma)$,
respectively (1ppb = $10^{-9}$).
Observations in the Milky Way have shown, however, a tentative signal
\dmm\ = $26 \pm 3$ ppb (Levshakov \etal\ 2010a,b), but its nature
remains unclear. 
To distinguish whether this is a real signal or an artefact due to unaccounted systematic
effects, additional independent measurements involving other molecules are required. 

One of such molecules~--- \chhhoh~--- was recently suggested 
by Jansen \etal\ (2011, hereafter J11).
\chhhoh\ is a widespread interstellar 
molecule observed in the Milky Way, external galaxies 
(Herbst \& van Dishoeck 2009; Sjouwerman \etal\ 2010; Mart\'in \etal\ 2006),
and even in comets (Bockel\'ee-Morvan \etal\ 1991).
The purpose of the present paper is to probe the variability of $\mu$ in the Milky Way 
using narrow emission lines of the methanol masers. 

The microwave spectrum of \chhhoh\ is very rich because of the internal rotation
of the OH group. The basic theory of the non-rigid tops with internal rotation
was established in the fifties (Lin \& Swalen 1959; Herschbach 1959)
and the main features of the methanol spectrum were explained. 
Later on the theory was refined many times and
currently there is a very impressive agreement between the theory and experiment
(Anderson \etal\ 1990; M\"uller \etal\ 2006; Xu \etal\ 2008; Kleiner 2010).

\section{Effective Hamiltonian and sensitivity coefficients}
\label{sect-2}

In this section the sensitivity coefficients $Q_\mu$ of the
methanol lines are calculated. Our computational procedure differs
from that described in J11. This allows us to check the values of $Q_\mu$
and their validity. 

To calculate the microwave spectrum of \chhhoh\
we use an approach from Rabli \& Flower (2010, hereafter RF), where a simple and
convenient form of the effective Hamiltonian 
with six spectroscopic constants is suggested. 
This Hamiltonian is physically transparent and sufficiently accurate for 
calculations of sensitivity coefficients. 
All six parameters of the RF model have clear physical meaning and their
dependence on $\mu$ is easily understood within the Born-Oppenheimer approximation.

To show how rotational parameters scale with $\mu$, we consider an example of 
a diatomic molecule in its ground vibrational state:
\begin{equation} \label{B0_scaling}
 B_0 = \frac{1}{M} \left\langle
 v=0\left|\frac{1}{R^2}\right|v=0\right\rangle
 =B_e -\alpha_e+\dots\, ,
\end{equation}
where $B_e=1/(MR_0^2)$ corresponds to the equilibrium internuclear distance
$R_0$ and $\alpha_e$ is the vibrational correction. It is clear that $B_e \sim
\mu$, but $\alpha_e$ has an additional dependence on $\mu$ via the vibrational wave
function and, hence, scales as $\mu^{3/2}$. 
Thus, there is no one-to-one correspondence between terms of the
effective Hamiltonian and the terms of the Born-Oppenheimer perturbation theory. 
As a result, $B_0$ scales as $\delta
B_0/B_0=Q^r_\mu \delta\mu/\mu$, with $Q^r_\mu \approx 1$. Typically,
$\alpha_e/B_0$ is of the order of $10^{-2}$ and $Q^r_\mu$ varies from 0.995
for the NO molecule to 0.981 for the H-bearing HF molecule.

We conclude that, in general, the rotational parameters $A$,
$B$, and $C$ scale linearly with $\mu$ within the uncertainty interval of 1--2\%.
These vibrational corrections are of
the same order of magnitude as centrifugal corrections considered in J11. 
Both types of corrections are
included below in the estimate of the error of the calculated sensitivity
coefficients.

The rotational part of the Hamiltonian $H_\mathrm{rot}$ corresponds to the
slightly asymmetric top and includes rotational constants $A$, $B$, and $C$
($B \approx C$). 
Here we use the standard convention $A > B > C$, while in RF $C > B > A$. 
The hindered rotation is described by the Hamiltonian
\begin{equation}\label{H1}
    H_\mathrm{hr}=-F\frac{{d}^2}{{d}\omega^2}
    + \frac{V_3}{2}\left(1-\cos\,3\omega \right)\,,
\end{equation}
where the kinetic coefficient $F$ is proportional to $\mu$ and the electronic
potential $V_3$ is independent on $\mu$. 
This model does not include
centrifugal distortions. Interaction of internal rotation with overall
rotation is described by a single parameter $D$, which scales linearly with
$\mu$ (Lin \& Swalen 1959). 
For the rotational degrees of freedom we use the basis set of
the prolate symmetric top and plane waves $\exp(im\omega)$ for the internal
rotation. The Hamiltonian (\ref{H1}) mixes waves with $m'-m=3n$. All relevant
matrix elements are tabulated in RF. The fitted values of the
parameters are also given there.

After the effective Hamiltonian is formed it is diagonalized numerically. Due to
the C$_3$ symmetry of the Hamiltonian (\ref{H1}) the final eigenstates can be
classified as A-type states and twofold degenerate E-type states. A-type
states have definite parity $p$: for A$^+$ states $p=(-1)^J$ and for A$^-$
states $p=(-1)^{J+1}$. Because methanol is close to the symmetric top its
states are classified with an approximate quantum number $K$, which
corresponds to the projection of the angular momentum $\bm{J}$ on the axis of
the CH$_3$ group. In this study we are interested only in the lowest states of
the internal rotation of A and E symmetry: excitation of higher states 
requires kinetic
temperatures $T \ga 300$K, while warm molecular clouds have typical
temperatures $T \la 100$K.

The \chhhoh\ transitions with $\Delta K=0$ can be considered approximately as rotational, 
where the
state of the internal motion does not change. These transitions have `normal'
sensitivities to $\mu$-variation with $Q_\mu \approx 1$. 
Because our model does not include centrifugal corrections, $Q_\mu$
must be exactly equal to 1. We estimate that these corrections
do not exceed 3\%\ for $J \le 10$. 
The J11 model accounts for these
corrections and provides the sensitivity coefficients for the 
$\Delta K=0$ transitions between 1.00 and 1.03 in agreement with our estimates.   
We note that this deviation from unity is of the same order of magnitude as
the vibrational correction which is not accounted for in both the RF and J11
effective Hamiltonians.  
The transitions with
$\Delta K=\pm 1$, on the contrary, lead to the change in the internal motion and
have sensitivities to $\mu$-variation which vary in a wide range. Such
transitions are of primary interest for our purpose.

In order to determine the sensitivity coefficients $Q_\mu$ for the microwave
transitions we first find the dependence of the eigenvalues $E_i$ on \dmm:
\begin{equation}\label{q1}
\Delta E_i = q_i \frac{\Delta \mu}{\mu}\ ,
\end{equation}
where the coefficient $q_i$, individual for each level, shows a response of the level $E_i$ 
to a small change of $\mu$ ($|\Delta \mu/\mu| \ll 1$). 
This is done by diagonalizing the effective
Hamiltonian for three sets of the parameters, which correspond to
$\mu=\mu_0$, and $\mu = \mu_0(1 \pm \veps)$, where $\veps$ equals to 0.001 or 0.0001
(in both cases we obtain the same $q$-factors). 
The parameters $A$, $B$, $C$, $D$, and $F$, all scaling linearly with $\mu$,
should then take the values $A_0$, $A_0 (1\pm \veps)$, etc., where the first
value $A_0$ corresponds to the fit of the experimental spectrum.
The dimensionless sensitivity coefficient $Q_\mu$ for the transition 
$\omega = E_{\rm up} - E_{\rm low}$
is found through calculation of the corresponding $q$-factor
\begin{equation}\label{q2}
\Delta \omega = q \frac{\Delta \mu}{\mu}\ ,
\end{equation}
where $ q = q_{\rm up} - q_{\rm low}$, and
\begin{equation}\label{Qmu}
    Q_\mu = \frac{q}{\omega}\ .
\end{equation}

The results of calculations for transitions with 
$\Delta K=\pm 1$ are listed in Table~\ref{tab_num},
where we also give experimental and calculated transition frequencies. One can
see that the accuracy of our model is about 1 GHz for all considered
transitions. Consequently, the relative error for the high frequency
transitions is only a fraction of a percent, but for the lowest frequency 6.7
GHz it is 15\%. In order to improve the accuracy of the 
$Q_\mu$-values we use experimental frequencies in Eq.~(\ref{Qmu}).
To check the computational procedure we also determined $Q_\mu$ for a few
transitions with $\Delta K = 0$ 
($0_0-1_0A^+$ 48.4 GHz, $1_0-2_0A^+$ 96.9 GHz, $1_1-2_1A^+$ 96.1 GHz, $1_1-2_1A^-$ 97.7 GHz, and 
$4_2-5_2A^-$ 242.2 GHz) 
and for all of them found $Q_\mu  = 1.00$.

Table~\ref{tab_num} also lists the errors of the sensitivity coefficients in their last digits
which are given in parenthesis. These errors were estimated in the following way. 
The effective Hamiltonian from RF 
includes the rotational and tunneling parts as well as the interaction between them.
The transition frequencies and the sensitivity coefficients are found from 
the numerical diagonalization of this
Hamiltonian followed by the numerical differentiation of the results in respect to $\mu$.

Only the low frequency transitions can have an enhanced sensitivity to the
$\mu$-variation because they are of a mixed character with the
rotational and tunneling contributions to the transition energy:
 \begin{equation}\label{err1}
 \omega = \omega_r - \omega_t\, .
 \end{equation}
These two contributions have the following dependences on $\mu$:
 \begin{equation}\label{err2}
 \frac{\Delta \omega_r}{\omega_r} = Q^r_\mu\frac{\Delta\mu}{\mu}\, ,
 \quad
 \frac{\Delta \omega_t}{\omega_t} = Q^t_\mu\frac{\Delta\mu}{\mu}\,.
 \end{equation}
Then, the resultant sensitivity of the transition $\omega$ is given by:
 \begin{equation}\label{err3}
 \frac{\Delta \omega}{\omega}
  =\left(Q^r_\mu\frac{\omega_r}{\omega}
  -Q^t_\mu\frac{\omega_t}{\omega}
  \right)\frac{\Delta\mu}{\mu}
  \equiv Q_\mu\frac{\Delta\mu}{\mu}\,.
 \end{equation}

Let us suppose that the sensitivities $Q^r_\mu$ and $Q^t_\mu$ 
are known with a relative error $\veps$. 
Then, the absolute error of $Q_\mu$ is equal to:
 \begin{equation}\label{err4}
 \Delta Q_\mu
  =\veps
  \times
  \frac{Q^r_\mu\omega_r +Q^t_\mu\omega_t}{\omega}\,.
 \end{equation}

In order to use this expression to estimate the errors 
$\Delta Q_\mu$ we need to know the decomposition (\ref{err1}) and the sensitivity coefficients
$Q^r_\mu$ and $Q^t_\mu$. As we discussed earlier, $Q^r_\mu\approx 1$. We can estimate
the tunneling sensitivity to be $Q^t_\mu\approx 2.6$  
from the semi-classical (WKB) approximation (e.g., Kozlov \etal\ 2010).
The tunneling part of the transition energy $\omega_t$ can be estimated from the
model of Hecht \& Dennison (1957).
The rotational energy $\omega_r$ is then given by the experimental
frequency $\omega$ and Eq.~(\ref{err1}).

We conservatively estimate the relative error of our calculations of $Q^r_\mu$
and $Q^t_\mu$ to be of about 3\%, i.e., $\veps=0.03$. This error is associated
with the missing centrifugal distortion and vibration correction
which changes the rotational and tunneling frequencies
by a few percent at the most. 
The resulting errors $\Delta Q_\mu$ are estimated from Eq.~(\ref{err4}).

Table~\ref{tab_num} shows that the sensitivity coefficients of 
the mixed rotation-tunneling transitions at low frequencies 
lie in the range between $-17(1)$ and $+43(3)$. For comparison,
the H$_2$ Lyman and Werner transitions have $Q_\mu \sim 0.03$ 
(Varshalovich \& Levshakov 1993), and 
the ammonia inversion transition has $Q_\mu = 4.46$ (Flambaum \& Kozlov 2007).
This means that methanol is almost 1000 times 
more sensitive to the change in $\mu$ than molecular hydrogen and about 10 times more
sensitive than ammonia.
Besides, \chhhoh\ has many lines with different sensitivities of both signs 
which allows us to estimate \dmm\ from observations of only this one molecule.
This is the advantage over, e.g., ammonia NH$_3$ which has the same sensitivity
coefficients for 
the inversion transitions (1,1), (2,2), etc., and, hence, requires some 
rotational transitions of other 
molecule as a reference in order to trace \dmm.

\begin{figure}[t]
\epsscale{2.0}
\plotone{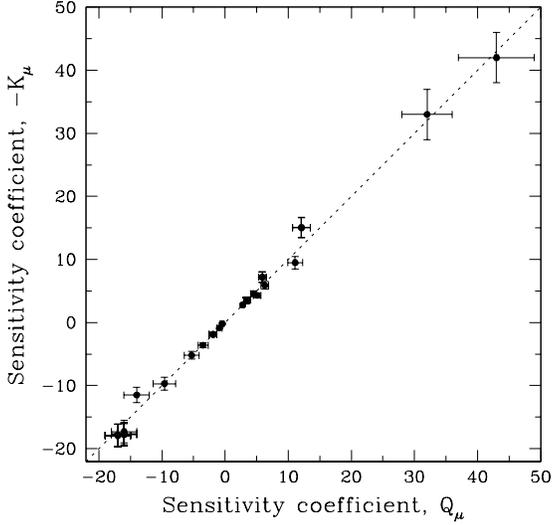}
\caption{Comparison of the sensitivity coefficients calculated in
the present paper ($Q_\mu$) with those from J11 ($-K_\mu$).
Shown by error bars are the $2\sigma$ uncertainties.
\label{fig1}}
\end{figure}

Now we can compare our $Q_\mu$ values with the sensitivity coefficients $K_\mu$ 
derived in J11. We note that $\mu$ was defined in J11 as $m_{\rm p}/m_{\rm e}$ and,
hence, one expects $Q_\mu = -K_\mu$. This comparison is shown graphically in Fig.~\ref{fig1}. 
The error bars in this figure mark the $1\sigma$ uncertainties.
In total we have 35 common computations of the sensitivity coefficients. 
The theoretical calculations of their values show a good concordance except for the six points
at 9.936, 37.703, 38.293, 38.453, 86.615, and 86.902 GHz where 
the discrepancy $|Q_\mu + K_\mu|$ 
exceeds the $2\sigma$ level (in Fig.~\ref{fig1}, the corresponding $Q_\mu$ values are
-14, 5.1, 12.1, 12.1, 5.9, and 5.9). 
The reason for such deviations is not clear. The highest offsets of $2.8\sigma$ and $2.7\sigma$
are found for the $J = 5-6$ transitions at 38.293 and 38.452 GHz, respectively, 
where the RF model should have a sufficiently accurate result.
On the other hand, the errors of the $K_\mu$ values were
taken in J11 to be 5\%\ if $|K_\mu| \geq 1$ (relative error) 
or 0.05 if $|K_\mu| < 1$ (absolute error).
However, analysis 
based on Eq.~(\ref{err4}) shows that the relative errors noticeably differ from line to line.

\begin{table*}[htb]
\begin{center}
\caption{Numerical calculation of the $Q$-factors for the low frequency mixed
rotation-tunneling transitions ($\Delta K=\pm 1$) in methanol. 
The rest frequencies are taken from M\"uller \etal\ (2004)$^\dag$
except for those marked by the asterisk which are from Lovas (2004).
The $1\sigma$ uncertainties of the least significant figure
of the rest frequencies and $Q_\mu$-values are given in parenthesis. 
}
\label{tab_num}
\begin{tabular}{r@{$-$}l r@{.}l r@{.}l c |  r@{$-$}l r@{.}l r@{.}l c}
\tableline\tableline
 \multicolumn{2}{c}{Transition} & \multicolumn{4}{c}{$\omega$\ (MHz)}&\multicolumn{1}{c}{$Q_\mu$} &
 \multicolumn{2}{|c}{Transition} & \multicolumn{4}{c}{$\omega$\ (MHz)}&\multicolumn{1}{c}{$Q_\mu$} \\
\multicolumn{2}{c}{ ${J_l}_{K_l} - {J_u}_{K_u}$} &\multicolumn{2}{c}{Exper.} & \multicolumn{2}{c}{Theor.} & 
\multicolumn{1}{c}{ } &
\multicolumn{2}{|c}{ ${J_l}_{K_l} - {J_u}_{K_u}$} &\multicolumn{2}{c}{Exper.} & \multicolumn{2}{c}{Theor.} \\[1pt]
\tableline
$6_0$& $5_1\ A^+$ & 6668&5192(4) & 5777&8 & +43(3) & $4_0$&$5_{-1}\ E $& 84521&169(10) & 85413&3 & $-3.5(4)$\\
$8_{-2}$ & $ 9_{-1}\ E$ & 9936&202(2) & 11281&9 & $-14(1)$ & $6_3$ & $7_2\ A^-$& 86615&600(5) & 86033&5 & +5.9(3)\\
$3_{-1}$ & $2_0\ E $& 12178&597(2) & 11487&4 &+32(2) & $6_3$ & $7_2\ A^+$& 86902&949(5) & 86315&2 & +5.9(3)\\
$3_0$ & $2_1\ E $& 19967&3961(2) & 20012&6 & +6.3(3) & $7_1$ & $8_0\ A^+$& 95169&463(10) & 96244&8 & $-1.9(3)$\\
$3_1$ & $3_2\ E $& 24928&707(7) & 25565&7 &$-17(1)$ & $10_{-2}$ & $11_{-1}\ E$&104300&414(7)&105949&0 & $-0.45(16)$\\ 
$4_1$ & $4_2\ E $& 24933&468(2) & 25597&1 &$-17(1)$ & $4_0$ & $3_1\ A^+$& 107013&803(5) & 106283&3 & +3.6(2)\\
$2_1$ & $2_2\ E $& 24934&382(5) & 25551&1 &$-17(1)$ & $1_{-1}$ & $0_0\ E $& 108893&963(7) & 108370&8 & +4.5(2)\\
$5_1$ & $5_2\ E $& 24959&0789(4)& 25655&3 &$-17(1)$ & $5_0$ & $6_{-1}\ E $& 132890&692(10) & 133898&0 & $-1.9(3)$\\
$6_1$ & $6_2\ E $& 25018&1225(4)& 25752&6 &$-16(1)$ & $8_1$ & $9_0\ A^+$& 146618&794(50) & 147799&4 & $-0.9(2)$\\
$7_1$ & $7_2\ E $& 25124&8719(4)& 25902&7 &$-16(1)$ & $8_{-1}$ & $8_0\ E $& 156488&868(10) & 155888&7 & +3.4(2)\\
$8_1$ & $8_2\ E $& 25294&4165(2)& 26120&3 &$-16(1)$ & $3_0$ & $2_1\ A^+$& 156602&413(10) & 155946&7 & +2.8(1)\\
$9_1$ & $9_2\ E $& 25541&3979(4)& 26419&4 &$-16(1)$ & $7_{-1}$ & $7_0\ E $& 156828&533(10) & 156258&3 & +3.4(2)\\
$3_1$ & $4_0\ E $& 28316&031(8)$^\ast$ & 28351&5 &$-2.8(3)$ & $6_{-1}$ & $6_0\ E$&157048&625(10)&156506&3 & +3.4(2)\\
$9_1$ & $8_2\ A^-$& 28969&942(50) & 29091&1 &+11.1(6)&$5_{-1}$ & $5_0\ E $& 157179&017(10) & 156661&9 & +3.4(2)\\
$3_0$ & $4_{-1}\ E $& 36169&265(30) & 36956&3 & $-9.6(9)$ &$4_{-1}$ & $4_0\ E $& 157246&056(10) & 156750&9 & +3.4(2)\\
$8_{-1}$ & $7_{-2}\ E $& 37703&700(30) & 36486&0 & +5.1(3) & $1_{-1}$ & $1_0\ E $& 157270&851(10) & 156817&2 & +3.4(2)\\
$5_3$ & $6_2\ A^-$& 38293&268(50) & 37660&7&+12.1(7) & $3_{-1}$ & $3_0\ E $& 157272&369(10) & 156795&3 & +3.4(2)\\
$5_3$ & $6_2\ A^+$& 38452&677(50) & 37817&3&+12.1(7) & $2_0$ & $2_1\ A^+$& 304208&324(13)$^\ast$ & 303751&3 & +1.91(7)\\
$6_1$ & $7_0\ A^+$& 44069&410(10) &45048&7& $-5.3(6)$ & $4_0$ & $4_1\ A^+$& 307165&911(13)$^\ast$ &306658&4& +1.89(7) \\
\tableline
\multicolumn{14}{l}{\footnotesize 
Note.~--- $^\dag$The uncertainties of the first three transitions 
at 6.668, 9.936, and 12.178 GHz
given in M\"uller \etal\ (2004)}\\[-1pt]
\multicolumn{14}{l}{\footnotesize 
correspond to the $2\sigma$ errors in accord with Breckenridge \& Kukolich (1995).}
\end{tabular}
\end{center}
\end{table*}

\section{Observational constraints on \dmm}
\label{sect-3}

The agreement between the values determined with
independent methods gives us confidence that, given
the high $Q_\mu$ values with different signs,
methanol is excellent for testing the electron-to-proton
mass ratio. 
Below we consider such a test based on high angular and
high spectral resolution observations of \chhhoh\ maser lines.

Equations~(\ref{q2}) and (\ref{Qmu}) show that
for a given transition from Table~\ref{tab_num},
$\omega_i$, with the sensitivity coefficient $Q_i$, the expected
frequency shift, $\Delta \omega_i/\omega_i$, due to a change in
$\mu$ is given by 
\begin{equation}\label{disEq1}
 \frac{\Delta \omega_i}{\omega_i} = Q_i\frac{\Delta \mu}{\mu}\, .
\end{equation}

Then the value of \dmm\ can be estimated 
from two transitions with different sensitivity coefficients $Q_i$ and $Q_j$:
\begin{equation}\label{disEq2}
 \frac{\Delta \mu}{\mu} = \frac{V_j - V_i}{c(Q_i - Q_j)}\, ,
\end{equation}
where $V_j$ and $V_i$ are the apparent radial velocities of the
corresponding \chhhoh\ transitions and $c$ is the speed of light.

Interstellar \chhhoh\ lines were widely observed in the last two decades
and we can obtain some preliminary estimates of \dmm\ using the published data.
In particular, the maser \chhhoh\ emission is of a special interest here since
maser lines are narrow and the error of the line center measurement is
correspondingly low.

The \chhhoh\ molecules originate in 
star-forming regions and are observed as maser emission in
two types of sources: class~I and class~II (e.g., Menten 1991). 
Class~II methanol masers are radiatively pumped and located in the vicinity
of young stellar objects (YSOs), whereas class~I methanol masers are believed to trace
distant parts of the outflows from YSOs and are collisionally pumped.
The most accurate \dmm\ values can be estimated from 
narrow and symmetric line profiles with similar shapes. 
Such profiles are provided in interferometric observations of the class~I 
masers located in the vicinity of
IRAS 16547--4247 (G343.12--0.06) which is a luminous YSO with a radio jet 
(Voronkov \etal\ 2006, hereafter V06). This methanol maser emission consists of
a cluster of six spots spread over an area of 30 arcsec in extent.
One spot (called $B$ in V06)
shows activity in 12 \chhhoh\ transitions (Fig.~3 of V06).
Among them, the 9.9 GHz and 104 GHz lines 
have a narrow spike ($FWHM < 30$ \ms) on top of a broader
($FWHM \approx 300$ \ms) symmetric line. The widths of these spikes do not exceed
the channel spacing 
which is 29 \ms\ and 22 \ms\ at 9.9 GHz and 104 GHz,
respectively. 
According to V06, these are
the narrowest spectral features ever found. Their linewidths 
imply that the masers are unsaturated and that the turbulent motion in the gas is
strongly suppressed.
The 9.9 GHz and 104 GHz masers show similar angular sizes, $\theta_{9.9} = 0.10\pm0.09$ arcsec,
and $\theta_{104} = 0.2\pm0.1$ arcsec, and almost the same coordinates, 
$|\Delta \alpha| = 0.03\pm0.04$ arcsec, and $|\Delta \delta| = 0.09\pm0.04$ arcsec.
The lines were observed in a month interval on June 16, 2005 (9.9 GHz) and on
Aug 18, 2005 (104 GHz).

As noted in V06, the 9.9 GHz and 104 GHz transitions belong to the same $J_{-2}-(J+1)_{-1}\ E$
transition series with $J = 8$ and 10, respectively, and are expected to show a similar behavior,
i.e., both maser lines should 
originate in the same volume and, hence,
have equal radial velocities. 
The radial velocities of these spikes, $V_{9.9} = -31.554$ \kms\ and $V_{104} = -31.594$ \kms,
are measured with the uncertainty of a few \ms, but, unfortunately, their difference, 
$\Delta V = V_{9.9} - V_{104} = 40$ \ms, is
less certain because of the errors in the rest frequencies, $\varepsilon_{9.9} = 60$ \ms\, and
$\varepsilon_{104} = 20$ \ms\ (see Table~\ref{tab_num}).
Since at these two frequencies the sensitivity coefficients 
are slightly different for two calculations
[$Q_{\mu, 9.9} = -14(1)$, $-K_{\mu, 9.9} = -11.5(6)$,
and $Q_{\mu, 104} = -0.45(16)$, $-K_{\mu, 104} = -0.18(5)$], we use their 
average values $\bar{Q} = (Q_\mu - K_\mu)/2$ to estimate \dmm\ from Eq.~\ref{disEq2}.
Thus, with
$\bar{Q}_{\mu, 9.9} = -12.75(58)$, $\bar{Q}_{\mu, 104} = -0.32(8)$
and $\Delta V = 40(63)$ \ms\ we find
\dmm\ = $11\pm17$ ppb ($1\sigma$, c.l.)
or the upper limit $|\Delta \mu/\mu| < 28$ ppb. 
This single point estimate
is in line with a limit derived from 
the sample mean of ammonia observations in the Milky Way,
$\Delta \mu/\mu = 26\pm3$ ppb ($1\sigma$, c.l.), or 
$|\Delta \mu/\mu| < 29$ ppb
(Levshakov \etal\ 2010a). 

The reliability of the present estimate of \dmm\ can be further
improved if new laboratory frequencies of the methanol 9.9 GHz and 104 GHz transitions
will be determined with a higher accuracy. 
Since maser sources are in general variable in time,
additional gains can be obtained from
simultaneous observations of several methanol lines
in a way described, e.g., in Voronkov \etal\ (2011) 
where up to eight methanol maser transitions were observed simultaneously with
the Australia Telescope Compact Array (ATCA).
Unfortunately, the observed methanol profiles at 24, 25 GHz and 9.9 GHz
have close sensitivity coefficients 
with $\Delta Q \approx 3$ which is comparable to the ammonia method.
This does not allow us to improve the aforementioned upper limit on \dmm. 
Without new laboratory studies 
and improvements in astronomical observations,
which require substantial care to determine frequencies with an accuracy
better than $10^{-8}$, 
any further advances in exploring $\Delta \mu/\mu$ from methanol maser spectra
will be impossible.

\acknowledgments

We thank Maxim Voronkov for details on \chhhoh\ maser emission from the spot $B$
towards G343.12--0.06, Irina Agafonova for her comments on the manuscript, and
Christian Henkel for many helpful discussions.
The work has been supported by the grant No.~`SFB 676 Teilprojekt C4',
the RFBR grants No.~09-02-00352 and No.~11-02-00943,  and by 
the State Program `Leading Scientific Schools of Russian Federation' (grant NSh-3769.2010.2).

\end{document}